\documentclass{PoS}
\usepackage{subfigure}
\usepackage{graphicx}
\usepackage{amssymb}
\usepackage{epstopdf}

\title{Controlling Residual Chiral Symmetry Breaking in Domain Wall Fermion Simulations}

\ShortTitle{Controlling Domain Wall Fermion Chiral Symmetry Breaking}

\author{\speaker{Dwight Renfrew} \\
Department of Physics, Columbia University, New York, NY 10027, USA \\
E-mail: \email{nhc@phys.columbia.edu}}

\author{Thomas Blum\\
Department of Physics, University of Connecticut, Storrs-Mansfield, CT 06269, USA \\
E-mail: \email{tblum@phys.uconn.edu}}

\author{Norman Christ\\
Department of Physics, Columbia University, New York, NY 10027, USA \\
E-mail: \email{nhc@phys.columbia.edu}}

\author{Robert Mawhinney\\
Department of Physics, Columbia University, New York, NY 10027, USA \\
E-mail: \email{rdm@phys.columbia.edu}}

\author{Pavlos Vranas\\
Lawrence Livermore National Laboratory, Livermore, CA 94550, USA \\
E-mail: \email{vranas2@llnl.gov}}

\abstract{At stronger gauge-field couplings, the domain wall fermion (DWF) residual mass ($m_{\rm res}$), a measure of chiral symmetry breaking ($\chi_{SB}$), grows rapidly as it is largely due to near zero fermion eigenmodes of $\log(T)$, where $T$ is the 4D transfer matrix along the fifth dimension.  The number of these eigenmodes increase rapidly at strong coupling.  To suppress such near-zero eigenmodes, we have added to the DWF path integral a multiplicative weighting factor consisting of a ratio of determinants of Wilson-Dirac fermions having a chirally twisted mass with a large negative real component and a small imaginary chiral component.  Numerical results show that this weighting factor with an appropriate choice of twisted masses significantly suppresses $m_{\rm res}$ while allowing adequate topological tunneling.

This research utilized resources at the New York Center for Computational Sciences at Stony Brook University/Brookhaven National Laboratory which is supported by the U.S. Department, of Energy under Contract No. DE-AC02-98CH10886 and by the State of New York.}

\FullConference{The XXVI International Symposium on Lattice Field Theory\\
July 14 - July 18 2008\\
Williamsburg, Virginia USA}

\begin{document}

It is known that the unwanted near-zero eigenmodes of the Wilson Dirac operator can be suppressed by including a suitable multiplicative weighting factor in the lattice path integral.  Vranas~\cite{Vranas:1999rz,Vranas:2006zk} demonstrated that a determinant of Wilson-Dirac fermions with real mass equal to the domain wall height acts as such a weighting factor for domain wall fermions (DWF).  Fukaya~\cite{Fukaya:2006vs} demonstrated that a ratio of determinants Wilson Dirac fermions with a real numerator mass and a chirally-twisted denominator mass acts as such a weighting factor for overlap fermions.

Here, we have selected as a weighting factor for DWF a ratio of determinants of Wilson Dirac fermions where both numerator and denominator have chirally-twisted masses.  The first section reviews properties of $m_{\rm res}$ useful for interpreting the action of the weighting factor; the second section characterizes this weighting factor, and the third section presents numerical studies of this weighting factor.  Readers interested in our numerical results can skip to the last section.

Note that this report goes beyond the previous conference presentation.  At the time of the conference, numerical studies has just begun and the reported results were of limited scope.  At this time, numerical studies have progressed and more interesting results can be reported.
\vspace{-0.1in} \section{Relevant Properties of DWF}
\label{sn-dwf}
$\chi_{SB}$ in domain wall fermions (DWF) can be represented in an effective low-energy Lagrangian as an extra fermion mass term with mass $m_{\rm res}$, where $m_{\rm res}$ can be defined as a ratio of the matrix element of the axial current at the 5D mid-plane and the matrix element of pseudo-scalar density.  A useful approximation to $m_{\rm res}$ is the following~\cite{Antonio:2008zz}.
\begin{equation}
m_{\rm res} \approx R^4_e \rho(\lambda_c)\frac{e^{-\lambda_c L_s}}{L_s} + R^4_l \rho(0) \frac{1}{L_s}
\label{eq:mres}
\end{equation}
Here, $\rho(\lambda)$ is the density of eigenmodes of a 4D Hamiltonian, $H_4=-\log(T)$, where $T$, a 4D transfer matrix acting along the 5'th dimension, is a functional of the Wilson-Dirac operator with negative mass, $D_\mathcal{W}(-M)$~\cite{Furman:1994ky}.  The term exponential in $L_s$ arises from eigenmodes with eigenvalues $|\lambda|$ greater than $\lambda_c$ (mobility edge) and have extended 4D support and exponential decay in the 5'th dimension. This term is easily controlled by increasing $L_s$.

At larger $L_s$ and stronger gauge-field couplings, the second term proportional to $L_s^{-1}$ (the inverse term) is larger and of principal interest.  This term arises from eigenmodes with near-zero eigenvalues (in any case $|\lambda|<\lambda_c$) having localized 4D support. Note that $\rho(0)$ is used to represent the density of near-zero eigenmodes, not of eigenmodes exactly at 0.

It is believed that many (perhaps all) 4D-localized, near-zero eigenmodes arise from patches of enhanced roughness in the gauge fields that are localized to a few lattice spacings.  Such patches have been associated artifactually with the appearance or disappearance of instantons in Monte Carlo time.  Gauge-field roughness, and therefore $\rho(0)$ and $m_{\rm res}$, increases rapidly as couplings, $\beta$, becomes stronger, and increase particularly rapidly at couplings below the QCD transition.

There is of course a caveat:  Proper sampling of gauge-field topology requires that the MD trajectories tunnel between topological sectors, and such tunneling is always associated with sign-change of a Dirac eigenmode.  MD trajectories therefore must not be prevented from passing through gauge fields where one or more Dirac eigenvalues are very close to 0 (exactly 0 in the continuum).  The very-close-to-zero eigenmodes relevant to topological tunneling are also thought to be 4D-localized.
\vspace{-0.1in} \section{The Auxiliary Determinant} \vspace{-0.1in}
Arguably the ideal weighting factor to suppress $m_{\rm res}$ would be built from $H_4$, which is computationally inaccessible.  Thus, our and others' weighting factors are built from the computationally-inexpensive Hermitian Wilson-Dirac operator $\gamma_5 D_\mathcal{W}(-M) \;\;\; M>0$.  As the determinant of this operator is near zero if one or more eigenmodes are near zero, including the squared determinant as a multiplicative factor in the DWF path integral will under-weight corresponding regions of gauge-field phase space during Monte Carlo sampling, or equivalently, generate a repulsive force away from such regions during molecular dynamics (MD) evolutions.\label{sn-auxdet}

We point out that this use of the Hermitian Wilson-Dirac operator relies on the assumption:  the eigenmodes operator $\gamma_5 D_\mathcal{W}(-M)$ are `sufficiently similar' to the those of the operator $H_4$ so that suppressing the near-zero eigenmodes of the former operator also suppresses the near-zero eigenmodes of the latter operator.  Similarity is to be expected, because, in the case of a continuous, unbounded 5'th dimension, $H_4$ become the Hermitian Wilson-Dirac operator, and, when this two operator has a zero eigenmode, the Hermitian Wilson-Dirac operator does also.
\begin{figure}[tbp]
\centering
\subfigure[No Wilson fermions]{
\includegraphics[trim=0.5in 1.0in 0.6in 0.5in,clip=true,width=2in]{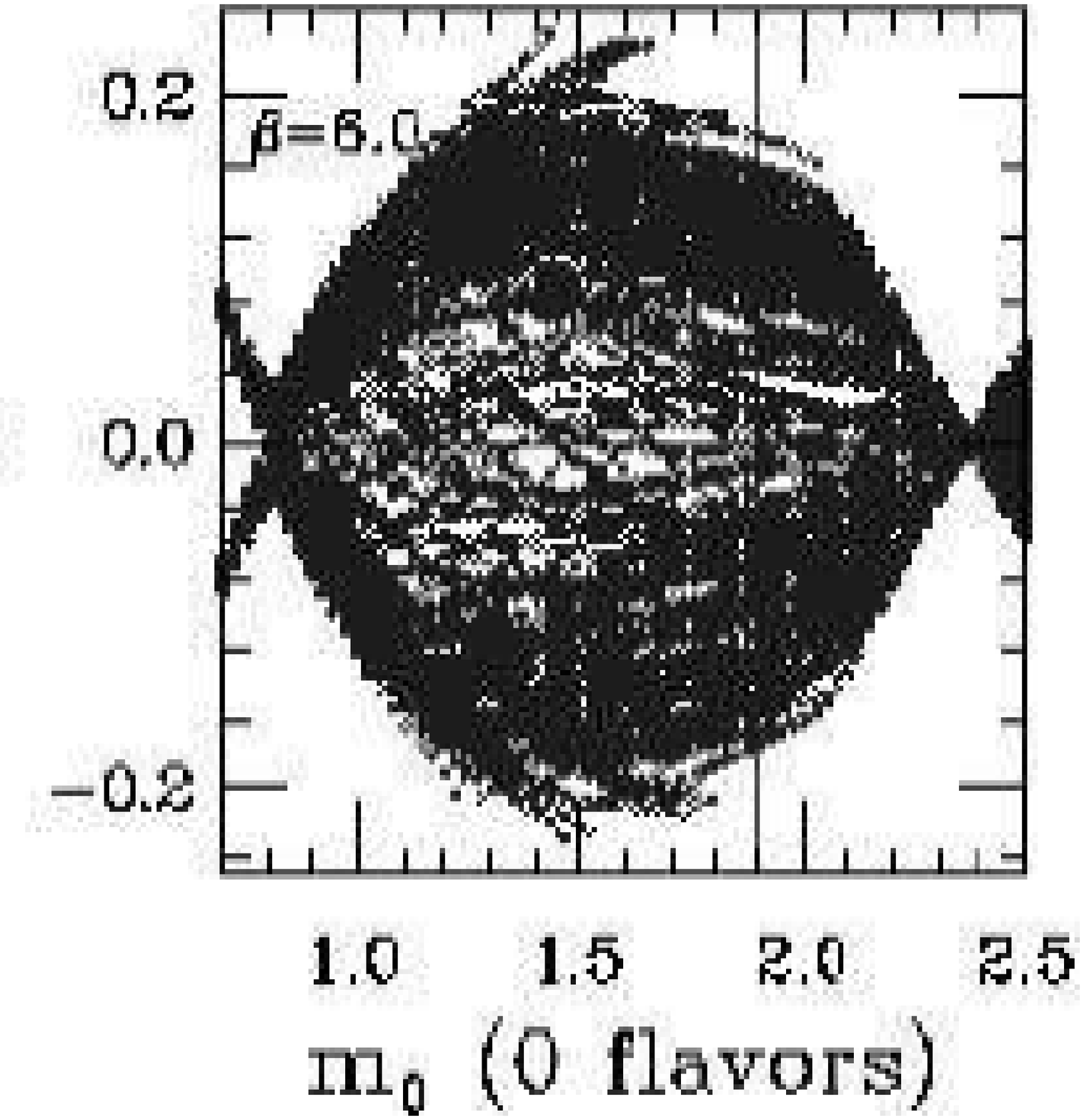}
\label{fg:vranas-b}
}
\subfigure[Two Wilson fermions]{
\includegraphics[trim=0.9in 1.05in 0.3in 0.5in,clip=true,width=2in]{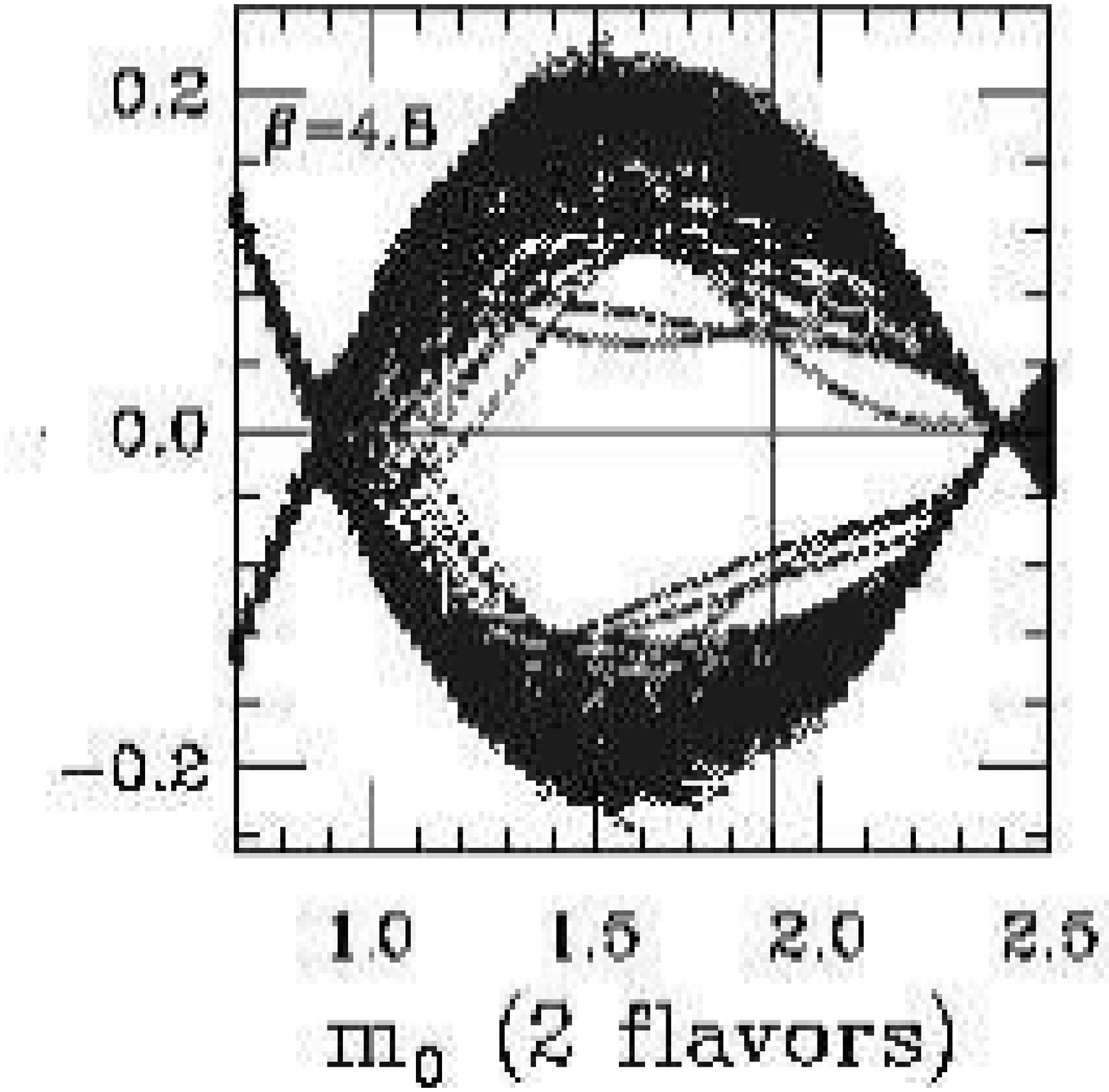}
\label{fg:vranas-a}
}
\label{fg:vranas}
\caption{Behavior of the 20 lowest eigenvalues of $D_\mathcal{W}(-m_0)$ versus $m_0$ at a lattice spacing of $a^{-1}\approx2.0$Gev for, (a), a pure Wilson gauge action, and, (b), a pure Wilson gauge action combined with two Wilson-Dirac fermions.}
\end{figure}
Vranas has provided a proof of principle that using a determinant of Wilson-Dirac fermions suppresses near-zero eigenmodes~\cite{Vranas:2006zk}.  Figure 1 from that paper demonstrates this suppression.

However, the Wilson fermion determinant is not an ideal weighting factor:  it essentially prevents  topological tunneling since it is very close to zero in the presence of very-close-to-zero eigenmodes; and it over-weights regions of gauge-field phase space where all eigenmodes are large since it is large in such regions.

Our weighting factor avoids these problems by including a small imaginary addition to the mass in the numerator (see refs.~\cite{Chen:2000zu} and~\cite{Vranas:2006zk}) and the denominator, the latter proposed in ref.~\cite{Fukaya:2006vs}.
\begin{eqnarray}
\mathcal{W}(M, \epsilon_f, \epsilon_b) =  \frac{\det[D_\mathcal{W}(-M+\imath\epsilon_b\gamma^5)^\dag D_\mathcal{W}(-M+\imath\epsilon_b\gamma^5)]}{\det[D_\mathcal{W}(-M+\imath\epsilon_f\gamma^5)^\dag D_\mathcal{W}(-M+\imath\epsilon_f\gamma^5)]} & \\
= \frac{\det[D_\mathcal{W}(-M)^\dag D_\mathcal{W}(-M)]+\epsilon_f^2}{\det[D_\mathcal{W}(-M)^\dag D_\mathcal{W}(-M)]+\epsilon_b^2}
= \prod_i \frac{\lambda^2_i+\epsilon^2_f}{\lambda^2_i+\epsilon^2_b} &
\end{eqnarray}
\begin{figure}[tbp]
\begin{center}
\includegraphics[trim=0.5in 1.0in 0.6in 0.5in,clip=true,width=2.5in,height=2.0in]{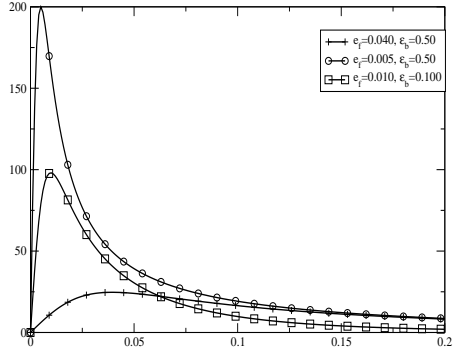}
\caption{$\mathcal{F}_i$ versus $|\lambda|$ for $\epsilon_f/\epsilon_b$ of $0.001/0.10$, $0.005/0.50$, and $0.040/0.50$.  Exemplary values at $|\lambda_i|$ of $0.001$, $0.025$, and $0.100$ of $\mathcal{F}_i(0.01,0.10)/\mathcal{F}_i(0.04,0.50)$ are: $16,\; 2.9$ and $0.6$; and of $\mathcal{F}_i(0.005,0.50)/\mathcal{F}_i(0.04,0.50)$ are $62,\; 3.4$ and $1.2$, respectively}
\label{fg-weight}
\end{center}
\end{figure}
Where $\lambda_i$ are the eigenvalues $ \gamma^5 D_\mathcal{W}$.  The effect of $\mathcal{W}$ can be more readily appreciated by considering the total force, $\mathcal{F}$, and the force from a single eigenmode, $\mathcal{F}_i$, generated during an MD simulation. \footnote{This expression for $\mathcal{F}$ abstracts features useful for the subsequent discussion.  A more correct derivation could proceed according to an \textit{ansatz} introduced in ~\cite{Gottlieb:1987mq}: first differentiate the MD Hamiltonian with respect to MD time, and then find the force as the traceless, anti-Hermitian (TA) part of the coefficient of the MD momentum. $\mathcal{F}$ is a formally color-independent coefficient and factors out of the TA operation. }
\begin{equation}
\mathcal{F}(\epsilon_f,\epsilon_b) \sim -\frac{\partial}{\partial \tau} (-\ln{\mathcal{W}}) \sim \sum_i \frac{d}{d \lambda_i }(-\ln{\mathcal{W}_i})\frac{\partial \lambda_i}{\partial U}\left(\frac{\partial U}{\partial \tau}\right)_{ii} \sim \sum_i \mathcal{F}_i(\epsilon_f,\epsilon_b) \; \frac{\partial \lambda_i}{\partial U}\left(\frac{\partial U}{\partial \tau}\right)_{ii}
\label{eq:fwi}
\end{equation}
$\tau$ is the MD time; $\mathcal{W}_i =  (\lambda^2_i+\epsilon^2_f)/(\lambda^2_i+\epsilon^2_b)$; and usually $0 < \epsilon_f^2 \ll \epsilon_b^2 < 1$.

We examine in detail the variation of $\mathcal{F}_i$ with $|\lambda_i|$, assuming that $(\partial \lambda_i/\partial U)$ varies sufficiently slowly with $|\lambda_i|$, for the three combinations of $\epsilon_f/\epsilon_b$ for which numerical results are presented below:  $0.001/0.10$, $0.005/0.50$, and $0.040/0.50$.  Figure \ref{fg-weight} illustrates the variation of $\mathcal{F}_i$ with $|\lambda_i|$ and lists exemplary values of the ratios $\mathcal{F}_i(0.01,0.10)/\mathcal{F}_i(0.04,0.50)$ and $\mathcal{F}_i(0.005,0.50)/$ $\mathcal{F}_i(0.04,0.50)$ for $|\lambda_i|$ of $0.001$, $0.025$, and $0.100$. This figure makes apparent that the size and position of the maximum of the force and the size and size and  position of the tail of the force can be significantly and \textit{independently} controlled by the imaginary mass parameters $\epsilon_f$ and $\epsilon_b$.  Table \ref{tb-weight} summarizes the $|\lambda_i|$-dependence of relative forces and eigenmode suppression.
\begin{table}
\caption{Dependence of relative forces and eigenmode suppression on $|\lambda_i|$.}
\label{tb-weight}
\begin{center}
\begin{tabular}{r p{4.5in}}
For $0.1 \lesssim |\lambda_i|$ & the forces satisfy $\mathcal{F}_i(0.005,0.5) \approx \mathcal{F}_i(0.04,0.5) \gtrsim 2 * \mathcal{F}_i(0.01,0.1)$; and accordingly the weighting factors $\mathcal{W}(0.005,0.5)$ and $\mathcal{W}(0.005,0.5)$ suppress eigenmodes in this range equally and to a greater extent than the weighting factor $\mathcal{W}(0.01,0.1)$. \\
For $|\lambda_i| \lesssim 0.025$ & the forces satisfy $\mathcal{F}_i(0.005,0.5) \gtrsim 2*\mathcal{F}_i(0.01,0.1) \gtrsim 4*\mathcal{F}_i(0.04,0.5)$; and accordingly the weighting factors suppress eigenmodes in this range in the order
$\mathcal{W}(0.005,0.5) \gg \mathcal{W}(0.01,0.1) \gg \mathcal{W}(0.04,0.5)$. \\
\end{tabular}
\end{center}
\end{table}
\vspace{-0.1in} \section{Numerical Studies}  \vspace{-0.1in}
The numerical results reported here have had the goal of rapidly obtaining a qualitative understanding of the properties of our selected weighting factor, and therefore have been continued only for sufficient trajectories, generally from 300 to 500 (MD timestep usually 0.2), to obtain, first, a reasonably accurate values for $m_{\rm res}$, second, an estimate the QCD transition temperature, $T_c$ ($\beta_c$), and third, an approximation of the topological charge.  Reference data without the weighting factor were drawn from prior studies of domain wall thermodynamics.  All simulations use the Iwasaki gauge action, have a weighting with two Wilson fermions in the numerator and two in the denominator, have two light DWF (masses: 0.03), and one heavy DWF (mass: 0.037) for $L_s=32$.  Further simulations details are the same as in~\cite{Antonio:2008zz} (except that this reference used $L_s=16$).

Physical scales have not, to date, been measured, nor has the susceptibility of $\bar{\Psi} \Psi$. Accordingly, we compare individual data values from different simulations only at the estimated $\beta_c$, and compare only groups of values above and below $\beta_c$.  Further, $\beta_c$ has been estimated by visual characteristics of the fluctuations of $\bar{\Psi}\Psi$ versus trajectory that were found to be reliable proxies for the susceptibility in prior thermodynamic studies.
\footnote{Two such characteristics are relied on.  First, in the vicinity of $\beta_c$, the amplitude of $\bar{\Psi}\Psi$, fluctuations is a maximum.  Second, the nature of these fluctuations differs above and below the transition.  For $\beta\gtrsim\beta_c$, $\bar{\Psi}\Psi$ has upward, short, relatively large-amplitude, fluctuations (spikes) without corresponding downward fluctuations superimposed on a largely constant baseline.  Spike amplitude and frequency increases as $\beta_c$ is approached.  For $\beta\lesssim\beta_c$, the baseline disappears and fluctuations become substantially symmetrical, and include low amplitude fluctuations extending over 10-50 trajectories.}
\begin{table}
\caption{Values of $m_{\rm res}$ for various $\epsilon_f$, $\epsilon_b$, and $\beta$.  On the left are thermodynamic runs for a series of $\beta$ at different combinations of $\epsilon_f$ and $\epsilon_b$. The horizontal double line through this sub-table identifies the estimated transition, and divides the table into an upper part where $\beta\lesssim\beta_c$ and a lower part where $\beta\gtrsim\beta_c$.  On the right are zero-temperature results for the series of $\epsilon_f$ at a selected $\beta$ and $\epsilon_b$. (* denote values that were determined from fewer than 300 trajectories.)}
\label{tb-mres}
\begin{center}
\begin{tabular}{||c|c||c|c||c|c||}
\hline
\multicolumn{6}{||c||}{\textit{lattice $16^3 \times 8 \times 32$}} \\
\hline
\multicolumn{2}{||c||}{\textit{no weighting factor}} &  \multicolumn{2}{|c||}{$\epsilon_f / \epsilon_b = 0.01/0.10$} &  \multicolumn{2}{|c||}{$\epsilon_f / \epsilon_b = 0.005/0.50$} \\
\hline
$\beta$ & $m_{\rm res}$ &  $\beta$ & $m_{\rm res}$ & $\beta$ & $m_{\rm res}$ \\
\hline
&           & 1.875 & 0.0101(5)* &  & \\
&           & 1.900 & 0.0072(3)* &  & \\
1.95 & 0.0252(1) & 1.925 & 0.0054(4)* &  & \\
2.00 & 0.0102(1) & 1.950 & 0.0035(3)* & 1.750 & 0.0015(1)   \\
\hline
\hline
2.05 & 0.0046(2) & 1.975 & 0.0026(4)* & 1.800 & 0.0009(3)* \\
2.08 & 0.0022(2) & 2.000 & 0.0020(2)* & 1.850 & 0.0003(4)*  \\
2.11 & 0.0011(1) &  &  &  &  \\
2.14 & 0.0007(1) &  &  &  &  \\
\hline
\end{tabular}
\hspace{0.15in}
\begin{tabular}{||c|c||}
\hline
\multicolumn{2}{||c||}{\textit{lattice $16^4 \times 32$}} \\
\hline
\multicolumn{2}{||c||}{$\beta=1.75 \;\; \epsilon_b=0.50$} \\
\hline
$\epsilon_f$ & $m_{\rm res}$ \\
\hline
&    \\
0.040 & 0.0025(1)*  \\
0.020 & 0.0018(1)   \\
0.005 & 0.0014(10)* \\
& \\
& \\
& \\
& \\
\hline
\end{tabular}
\end{center}
\end{table}
\begin{table}
\vspace{-0.1in}
\caption{Values of the average of the absolute topological charge, $<|\nu_{top}|>$, for various $\epsilon_f$, $\epsilon_b$, and $\beta$.  The left and right sub-tables have the same structure as in Table 1. Runs were too short to determine useful error bounds; it is believed that errors where $<|\nu_{top}|>=O(1)$ may be up to 50\%, and where $<|\nu_{top}|>=O(0.1)$ may be up to 100\%. (* denote values that were determined from fewer than 300 trajectories.)}
\label{tb-top}
\begin{center}
\begin{tabular}{||c|c||c|c||c|c||}
\hline
\multicolumn{6}{||c||}{\textit{lattice $16^3 \times 8 \times 32$}} \\
\hline
\multicolumn{2}{||c||}{\textit{no determinant}} &  \multicolumn{2}{|c||}{$\epsilon_f / \epsilon_b = 0.01/0.10$} &  \multicolumn{2}{|c||}{$\epsilon_f / \epsilon_b = 0.005/0.50$} \\
\hline
$\beta$ & $<|\nu_{top}|>$ &  $\beta$ & $<|\nu_{top}|>$ & $\beta$ & $<|\nu_{top}|>$ \\
\hline
&      & 1.875 & 1.6* &      &      \\
&      & 1.900 & 1.7* &      &      \\
1.950 & 2.7 & 1.925 & 1.0* &      &      \\
2.125 & 1.7 & 1.950 & 0.8* & 1.75 & 0.1    \\
\hline
\hline
2.050 & 0.9 & 1.975 & 0.5* & 1.80 & 0.2*    \\
&      & 2.000 & 0.4* & 1.85 & 0.0*  \\
\hline
\end{tabular}
\hspace{0.15in}
\begin{tabular}{||c|c||}
\hline
\multicolumn{2}{||c||}{\textit{lattice $16^4 \times 32$}} \\
\hline
\multicolumn{2}{||c||}{$\beta=1.75 \;\; \epsilon_b=0.50$} \\
\hline
$\epsilon_f$ & $<|\nu_{top}|>$ \\
\hline
&    \\
0.040 & 2.1* \\
0.020 & 0.1 \\
0.005 & 0.4* \\
& \\
& \\
\hline
\end{tabular}
\end{center}
\vspace{-0.1in}
\end{table}
We interpret the data in Tables \ref{tb-mres} and \ref{tb-top} in view of Figure \ref{fg-weight} and the discussion in Section \ref{sn-dwf}.  The data are consistent with the following hypothesis.
\begin{center}
\begin{tabular}{p{1.5in} p{4.0in}}
$m_{\rm res}$ suppression & Depends on suppression of near-zero eigenmodes of the Hermitian Wilson-Dirac operator having $|\lambda|$ values less than about $0.1-0.5$; \\
Topological charge suppression & Depends on suppression of very-close-to-zero eigenmodes of the Hermitian Wilson-Dirac operator having $|\lambda|$ values less than about $0.01-0.05$; \\
\label{tb:hypo}
\end{tabular}
\end{center}  \vspace{-0.2in}
First, compare the expected suppression of eigenmodes with $0.1 \lesssim |\lambda_i|$ from Table \ref{tb-weight} with the $m_{\rm res}$ measurements from Table \ref{tb-mres}.  It is expected that, in comparison to $m_{\rm res}$ without any weighting factor, $m_{\rm res}$ is most suppressed at $0.005/0.50$; suppressed to a somewhat less degree at $0.04/0.50$; and clearly less suppressed at $0.01/0.10$.  As expected, measured $m_{\rm res}$, in comparison to $m_{\rm res}$ without any weighting factor, is most suppressed at $0.005/0.50$ and less suppressed at $0.01/0.10$.  $m_{\rm res}$ suppression at $0.04/0.50$ is intermediate between that at $0.005/0.50$ and at $0.01/0.10$. \footnote{$m_{\rm res}$ is known to be largely independent of temperature, as is confirmed by comparing $m_{\rm res}$ at $\epsilon_f/\epsilon_b=0.005/0.50$ and $\beta=1.75$ from the zero temperature lattice $16^4 \times 32$ with that from the non-zero temperature lattice $16^3 \times 8 \times 32$.}

Further, comparing $m_{\rm res}$ without any weighting factor to $m_{\rm res}$ at $0.01/0.10$ above and below the transition: below the transition, relative suppression of $m_{\rm res}$ is greater than at the transition; and above the transition, relative suppression is less than at the transition.  Accordingly, this weighting factor can be expected to be increasingly effective as gauge-field coupling increases.

Next, compare the expected suppression of eigenmodes with $|\lambda_i| \lesssim 0.025$ from Table \ref{tb-weight} with the topological charge measurements from Table \ref{tb-top}.  It is expected that, in comparison to topological charge without any weighting factor, topological charge is most suppressed at $0.005/0.50$; less suppressed at $0.01/0.10$; and minimally suppressed at $0.04/0.50$.  Again as expected, measured topological charge, in comparison to topological charge without any weighting factor, is virtually eliminated at $0.005/0.50$ and suppressed to an intermediate degree at $0.01/0.10$.  Although data is limited, the topological charge at $0.04/0.50$ is likely to be closer to the topological charge without any weighting factor than to the topological charge at $0.01/0.10$.

We conclude that near-zero eigenmodes of the Hermitian Wilson Dirac operator can usefully serve as proxies for the near-zero eigenmodes of $H_4$, and also that the near-zero and very-close-to-zero eigenvalue ranges are reasonable for the characteristics of the DWF used here.  It is likely that the eigenvalue ranges that here appear relevant to suppression of $m_{\rm res}$ and of the topological charge are specific to the parameters defining the DWF fermions simulated here, particularly the explicit light quark masses. For example, in the case of a different explicit light quark mass, the eigenvalue ranges relevant to suppression of the topological charge can be expected to change\footnote{Noting that  $\det[D_{DWF}(m_f)] = 0.5*\det[D_{DWF}(1)]^{-1}\det\left[ (1+m_f)\mathbb{I} +(1-m_f)\Gamma_5\tanh(-(L_s/2)\log(T)\right]$, where T, the transfer matrix, is independent of $m_f$; and $\det[D_{DWF}(1)]$ is the Pauli-Villars subtraction.}. However, the eigenvalue ranges relevant to suppression of $m_{\rm res}$ can be expected to be little changed, as $H_4$ is independent of explicit quark mass.

Nevertheless, the parameters studied here may well be close to those needed to allow the use of DWF for larger lattice spacing and consequently on larger lattice volumes with $m_{\rm res}$ values better controlled that has been possible to date.  At present the largest physical volumes that have been accessible are $L \approx 2.7$fm boxes used with $24^3\times64$ lattices at $1/a=1.73$Gev and $32^3\times64$ lattices at $1/a=2.42$Gev work of RBC/UKQCD with $m_{\rm res}$ values of 0.00315(2) and 0.000676(11), respectively.  Pending a more accurate determination of the lattice scale, the results presented here suggest that DWF simulations with $m_{\rm res} \approx 0.002$ and $1/a=1.4$fm giving a $32^3$ lattice volume with a linear size of $L=4.5$fm may be possible.  Such larger physical lattice volumes will offer important opportunities for study of QCD thermodynamics for study of $K \rightarrow \pi\pi$ decays.

\bibliographystyle{JHEP}
\bibliography{lat}

\pagebreak
\begin{center}
{\Large \textbf{Appendix}}
\end{center}
\vspace{-0.1in}
Subsequent to submitting this report to the Proceedings of Science, we determined the behavior of several lowest Wilson-Dirac eigenvalues versus mass for sample configurations from the runs reported in Table \ref{tb-mres}, and present in this Appendix results for thermalized sample configurations from the three runs appearing above the double line in the left portion of  Table \ref{tb-mres}.\footnote{Diagrams of this type are often presented $-m_0\ge0$ on the horizontal axis.}.  These three runs are expected to be below, but near to, the QCD transition.  When no weighting factor is used, the transition has been independently determined to be at or near $\beta=2.031(5)$ \footnote{Michael Cheng et al., \textit{The transition temperature using 2+1 flavors of domain wall fermions at Nt = 8}, in preparation.}.

These results are sufficient to qualitatively appreciate the effects of the weighting factor.  However, the statistics are insufficient for quantitative analysis at this time.  In all cases, the domain wall fermion height was -1.8.
\begin{figure}[htb]
\begin{center}
\includegraphics[trim=0.5in 0.7in 0.6in 0.6in,clip=true,width=2.5in,height=2.0in]{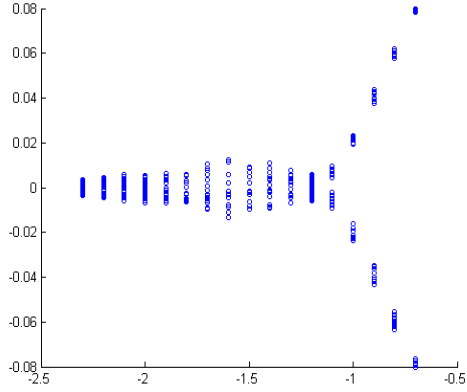}
\caption{Without weighting factor: behavior of the 12 lowest eigenvalues of $D_\mathcal{W}(m_0)$ versus $m_0$ at for a thermalized $16^3 \times 8 \times 32$ lattice at $\beta=1.950$.}
\label{fg-thermo200}
\end{center}
\end{figure}
\vspace{-0.1in}
\begin{figure}[htb]
\centering
\subfigure[$\beta=1.950$ and $\epsilon_f / \epsilon_b = 0.01/0.10$.]{
\includegraphics[trim=0.5in 0.7in 0.6in 0.68in,clip=true,width=2.75in]{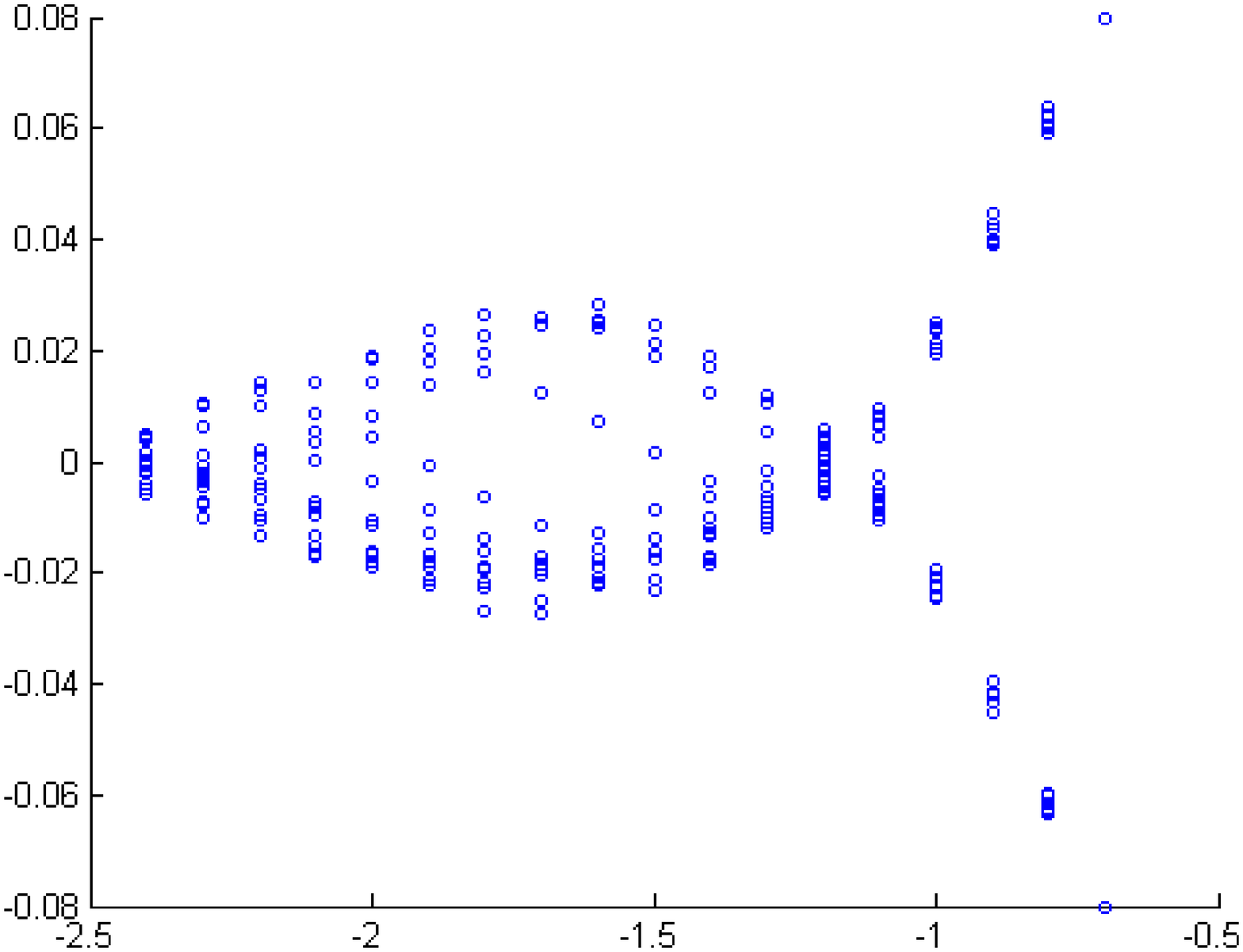}
\label{fg:vranas1}
}
\subfigure[$\beta=1.750$ and $\epsilon_f / \epsilon_b = 0.005/0.50$.]{
\includegraphics[trim=0.5in 0.7in 0.6in 0.68in,clip=true,width=2.75in]{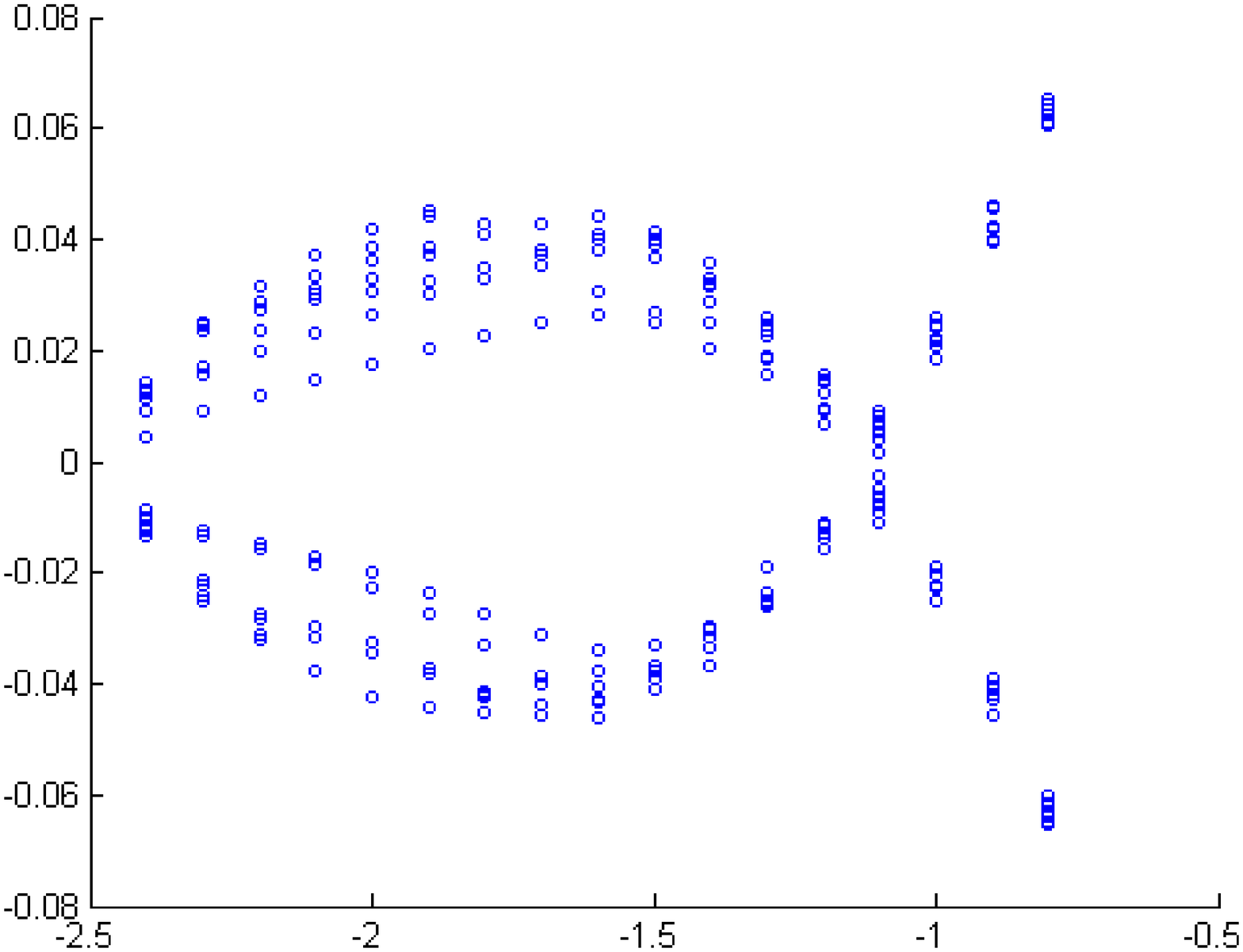}
\label{fg:vranas2}
}
\label{fg:vranas3}
\caption{With weighting factor: behavior of the 12 lowest eigenvalues of $D_\mathcal{W}(m_0)$ versus $m_0$ at for a thermalized $16^3 \times 8 \times 32$ lattice at $\beta=1.950$ and $\epsilon_f / \epsilon_b = 0.01/0.10$ (left subfigure) and for
 $\beta=1.750$ and $\epsilon_f / \epsilon_b = 0.005/0.50$ (right subfigure).}
\end{figure}
\end{document}